\documentclass[12pt]{article}
\usepackage{amsmath}
\usepackage{times}
\usepackage{graphicx}
\usepackage{color}
\usepackage{comment}
\usepackage{multirow}
\usepackage{amssymb}
\usepackage[round,authoryear]{natbib}
\usepackage{rotating}
\usepackage{latexsym}
\usepackage{setspace}

\newcommand{\captionfonts}{\normalsize}

\makeatletter  
\long\def\@makecaption#1#2{%
  \vskip\abovecaptionskip
  \sbox\@tempboxa{{\captionfonts #1: #2}}%
  \ifdim \wd\@tempboxa >\hsize
    {\captionfonts #1: #2\par}
  \else
    \hbox to\hsize{\hfil\box\@tempboxa\hfil}%
  \fi
  \vskip\belowcaptionskip}
\makeatother   

\begin{document}
\hspace{13.9cm}1

\ \vspace{20mm}\\

{\LARGE On the spike train variability characterized by variance-to-mean power relationship}

\ \\
{\bf \large Shinsuke Koyama}\\
{Department of Statistical Modeling, 
The Institute of Statistical Mathematics, 
Tokyo, Japan
}


\thispagestyle{empty}
\markboth{}{NC instructions}
\ \vspace{-0mm}\\
%
\begin{center} {\bf Abstract} \end{center}

We propose a statistical method for modeling the non-Poisson variability of spike trains observed in a wide range of brain regions. Central to our approach is the assumption that the variance and the mean of interspike intervals are related by a power function characterized by two parameters: the scale factor and exponent.
It is shown that this single assumption allows the variability of spike trains to have an arbitrary scale and various dependencies on the firing rate in the spike count statistics, as well as in the interval statistics, 
depending on the two parameters of the power function. 
We also propose a statistical model for spike trains that exhibits the variance-to-mean power relationship, and based on this a maximum likelihood method is developed for inferring the parameters from rate-modulated spike trains. The proposed method is illustrated on simulated and experimental spike trains.

\section{Introduction}

The variability of neural firing is of central importance in the study of signal processing that is carried out by the nervous system. 
The reliable transmission of sensory signals, integration of neural information, and precise control of neural-motor systems are significantly dependent on the variability of the neural responses to identical sensory or behavioral variables, as well as on the average responses \citep{Mainen95,deRuyter97,Harris98,Shadlen98,Ma06,Lu13}.
 
Two types of measurement, inter-spike interval (ISI) and spike count, are commonly used to quantify the variability of spike trains. 
The variability of ISI, expressed in the variance, quantifies how irregular the firing time is on a short timescale, characterized by the typical ISI. Since the variance of ISI is computed within single spike trains, it signifies intra-trial variability.
The variance of the spike count across repeated observations, by contrast, quantifies the trial-to-trial variability in relatively long time intervals.
These two quantities are by no means independent variables, but are closely related  \citep{Nawrot08}. 
In general, the variances of both ISI and spike count are scaled by the mean, the degree of which may vary across different brain regions \citep{Kara00,Maimon09}.

In this article, we formulate a statistical framework for modeling the variability of spike trains in terms of both the ISI and counting statistics. 
Our approach is motivated by an observation made by \cite{Troy92}. 
They reported that for steady discharges of X retinal ganglion cells of cats, in response to stationary visual patterns, the standard deviation of ISI increases as approximately the 3/2 power of the mean ISI.
Motivated by their observation, we make a single assumption about the ISI statistics:
\begin{equation}
\mathrm{Var(ISI)} = \phi\mathrm{E(ISI)}^{\alpha},
\label{eq:statlaw}
\end{equation}
where 
$\phi$ is the scale factor controlling the overall variance of ISIs, and $\alpha$ is the exponent controlling how the variance is scaled by the mean.
Presently, it should be emphasized that this statistical assumption is a generalization of the finding of \cite{Troy92}, in the sense that $\phi(>0)$ and $\alpha$ can take arbitrary values in theory. 
On the basis of the power law (\ref{eq:statlaw}), 
we show that this allows the spike trains to have a wide range of variability in the counting statistics, as well as in the ISI statistics observed across the brain areas, depending on $\phi$ and $\alpha$.
By combining Eq.~(\ref{eq:statlaw}) with the time-rescaling transformation \citep{Barbieri01}, we propose a ``generalized" rate-modulated renewal process to model spike trains, and develop a maximum likelihood method to infer $\phi$ and $\alpha$ from rate-modulated spike trains. 

The rest of this article is organized as follows. 
In section~\ref{sec:theory}, we develop a statistical method. 
In section~\ref{sec:result}, we illustrate our method on simulated and experimental data. 
Section~\ref{sec:discussion} contains discussions on the possible implications of the results.

\section{Theory} \label{sec:theory}
\subsection{Statistical assumption}

Consider spike trains whose ISIs are independent and identically distributed, with mean $\mu$ and variance $\sigma^2$.
The central assumption in our approach is that the variance of ISI has a power function of the mean, in the form
\begin{equation}
\sigma^2 = \phi \mu^{\alpha},
\label{eq:scaling-isi}
\end{equation}
where $\phi>0$ is the scale factor controlling the overall amplitude of the power law, and $\alpha$ is the exponent controlling how the variance is scaled by changing the mean.
For $\alpha=2$, the scale factor $\phi$ corresponds to the squared coefficient of variation, whose value is unity for a Poisson process. 
By contrast, values of $\alpha>2(<2)$ imply that the timing of spike tends to be over (under) dispersed for large means, and under (over) dispersed for small means.

Next, consider the spike count.  
Let $N_{\Delta}$ be the number of spikes in the counting window of duration $\Delta$. 
The variability of spike count is often measured by the Fano factor, defined by the ratio of the variance to the mean: 
\begin{equation}
F_{\Delta} := \frac{\mathrm{Var}(N_{\Delta})}{\mathrm{E}(N_{\Delta})},
\end{equation}
where the expectation is computed over repeated observations.
For a large counting window $\Delta\gg\mu$, the mean and variance of $N_{\Delta}$ are asymptotically evaluated as $\mathrm{E}(N_{\Delta})\sim\Delta/\mu$ and $\mathrm{Var}(N_{\Delta})\sim\sigma^2\Delta/\mu^3$, respectively \citep{Cox62}.
Suppose that the variance of ISIs obeys Eq.~(\ref{eq:scaling-isi}). 
Then, for large $\Delta$ the Fano factor exhibits the power law 
\begin{equation}
F_{\Delta} \sim \phi \lambda^{\gamma},
\label{eq:fano-asymptotic}
\end{equation}
where
\begin{equation}
\lambda := \frac{\mathrm{E}(N_{\Delta})}{\Delta}
\end{equation}
is the mean firing rate, and the exponent $\gamma$ is related to that of the ISI statistics via the scaling relation:
\begin{equation}
\gamma = 2-\alpha.
\label{eq:exponent}
\end{equation}
Eq.~(\ref{eq:fano-asymptotic}) describes the dependency of the Fano factor on the ISI parameters and the mean firing rate $\lambda$.
For $\gamma=0$ (i.e., $\alpha=2$), the Fano factor does not depend on $\lambda$; 
in other words, the variance of the spike count is proportional to the mean.
If $\gamma>0$ ($\alpha<2$), the Fano factor increases as $\lambda$ increases, while the Fano factor is inversely related to $\lambda$ if $\gamma<0$ ($\alpha>2$).

The Fano factor depends on the length of the counting window $\Delta$. 
When $\Delta\ll\mu$, the probability of two and more spikes is negligible, and the spike count can be approximated by a Bernoulli random variable with probabilities $P(N_{\Delta}=1)=\lambda\Delta$ and $P(N_{\Delta}=0)=1-\lambda\Delta$, respectively. 
The variance of the Bernoulli distribution is $\lambda\Delta(1-\lambda\Delta)$, so that for any values of $\alpha$ and $\phi$ the Fano factor approaches unity \citep{Teich97}:
\begin{equation}
\lim_{\Delta\to0}F_{\Delta} = \lim_{\Delta\to0}\frac{\lambda\Delta(1-\lambda\Delta)}{\lambda\Delta}
=1,
\label{eq:fs-small-delta}
\end{equation}
which is different from Eq.~(\ref{eq:fano-asymptotic}).
In the numerical studies presented in section~\ref{sec:result}, we choose $\Delta$ so that an average of five spikes fall in the window, which is enough for Eq.~(\ref{eq:fano-asymptotic}) to apply.

\subsection{Statistical model}
\subsubsection{Generalized rate-modulated renewal process}

We construct a statistical model for spike trains whose variability is characterized by 
the variance-to-mean power law. 
Consider first the stationary renewal process, 
a class of point processes in which ISIs are independent and identically distributed. 
Let $X$ be a random variable describing ISI.  
It follows from Eq.~(\ref{eq:scaling-isi}) that by rescaling ISI as $X\to \lambda X$, $\lambda=1/\mu$ being the mean firing rate, the parameters are rescaled as 
$\mu\to 1$ and $\phi\to \lambda^{2-\alpha}\phi$.
Thus, a parametric probability density function $f(x;\mu,\phi)$ that has mean $\mu$ and variance $\phi\mu^{\alpha}$, 
and is invariant under the rescaling, satisfies 
\begin{equation}
f(x;\mu,\phi) = \lambda f(\lambda x; \lambda^{2-\alpha}\phi),
\label{eq:scaling-pdf}
\end{equation}
where $f(x;\phi):=f(x;1,\phi)$.
Eq.~(\ref{eq:scaling-pdf}) suggests that 
one can always reparametrize an arbitrary probability density function with unit mean and variance $\phi$, so that the variance has the power function of the mean (\ref{eq:scaling-isi}).

We extend the stationary renewal process defined by Eq.~(\ref{eq:scaling-pdf}), to a rate-modulated process.
Let $N(t)$ be the number of spikes that have already occurred at time $t$, and $t_i$ denote the $i$th spike time. 
A point process is fully defined by a conditional intensity function \citep{Daley03,Kass-Ventura01}, 
\begin{equation}
r(t;H(t)) = \lim_{dt\to0} \frac{P\{N(t+dt)-N(t)=1;H(t)\}}{dt},
\label{eq:cif-general}
\end{equation}
where $H(t)=\{t_1,t_2,\ldots,t_{N(t)}\}$ denotes the history of spikes up to the time $t$. 
For a renewal process whose ISI density function is given by $f(x;\phi)$, the conditional intensity function, also called the hazard function, is given by 
\begin{equation}
r(t;t_{N(t)},\phi) = \frac{f(t-t_{N(t)};\phi)}{1-\int_{t_{N(t)}}^tf(u-t_{N(t)};\phi)du}. 
\label{eq:cif-renewal}
\end{equation}
Let $\lambda(t) >0$ be an instantaneous firing rate, and define 
\begin{equation}
\Lambda(t)=\int_0^t\lambda(u)du,
\end{equation} 
which is monotone and invertible.
By rescaling the time $t\to\Lambda(t)$, we can obtain the ``conventional" rate-modulated renewal process \citep{Barbieri01,Berman81,Koyama08,Koyama14,Nawrot08,Pillow08,Reich98}, whose conditional intensity function is given by
\begin{equation}
r(t;t_{N(t)},\{\lambda(t)\},\phi) = 
\frac{\lambda(t)f(\Lambda(t)-\Lambda(t_{N(t)});\phi)  }
{1-\int_{t_{N(t)}}^t\lambda(v) f(\Lambda(v)-\Lambda(t_{N(t)});\phi)dv}.
\label{eq:cif-ns-cnv}
\end{equation}
Note that the expectation of Eq.~(\ref{eq:cif-ns-cnv}) is equal to the following:
\begin{equation}
\lambda(t) = \mathrm{E}[r(t;t_{N(t)},\{\lambda(t)\},\phi)],
\end{equation}
where $\lambda(t)$ is also called the ``marginal" intensity function, which does not depend on the past spikes. 
However, the Fano factor of the process (\ref{eq:cif-ns-cnv}) does not have the power law with the exponent (\ref{eq:exponent}).\footnote{
In fact, this transformation results in the Fano factor being a constant, whose value is determined by $\phi$.
}

We generalize Eq.~(\ref{eq:cif-ns-cnv}) such that the Fano factor has a power function of the firing rate.
Analogously with Eq.~(\ref{eq:scaling-pdf}), by rescaling the parameter $\phi\to \lambda(t)^{2-\alpha}\phi$, as well as the time $t\to\Lambda(t)$, the conditional intensity function of a ``generalized" rate-modulated renewal process is obtained as 
\begin{equation}
r(t;t_{N(t)},\{\lambda(t)\},\phi,\alpha) = 
\frac{\lambda(t)f(\Lambda(t)-\Lambda(t_{N(t)});\lambda(t)^{2-\alpha}\phi)  }
{1-\int_{t_{N(t)}}^t\lambda(v) f(\Lambda(v)-\Lambda(t_{N(t)});\lambda(v)^{2-\alpha}\phi)dv}.
\label{eq:cif-ns}
\end{equation}
Eq.~(\ref{eq:cif-ns}) is reduced to the conditional intensity function associated with Eq.~(\ref{eq:scaling-pdf}) if $\lambda(t)=\lambda$, and corresponds to Eq.~(\ref{eq:cif-ns-cnv}) if $\alpha=2$.

\subsubsection{Likelihood function}
Using the conditional intensity function (\ref{eq:cif-ns}), 
the probability density of the spike trains $\{t_i\}:=\{t_1,t_2,\ldots,t_n\}$ in the interval $(0,T]$ can be expressed as 
\begin{eqnarray}
\lefteqn{p(\{t_i\};\{\lambda(t)\},\phi,\alpha)}\hspace{1cm}\nonumber\\
 &=& 
P_1(t_1)
\prod_{i=2}^n r(t_i;t_{i-1},\{\lambda(t)\},\phi,\alpha)  \nonumber\\
& & { } \times
\exp\Bigg[
-\int_{t_1}^{t_n} r(u;t_{N(u)},\{\lambda(t)\},\phi,\alpha) du
\Bigg] P_0((t_n,T]),
\label{eq:likelihood1}
\end{eqnarray}
where $P_1(t_1)$ is the probability of the first spike occurring at time $t_1$, 
$P_0((t_n,T])$ is the probability of no spikes occurring in the interval $(t_n,T]$, and 
the exponential factor represents the probability of there being no spikes in each interspike interval \citep{Daley03,Kass-Ventura01}. 
Substituting Eq.~(\ref{eq:cif-ns}) into Eq.~(\ref{eq:likelihood1}) yields the more tractable form (see Appendix~\ref{apdx:likelihood}):
\begin{eqnarray}
\lefteqn{p(\{t_i\};\{\lambda(t)\},\phi,\alpha)}\hspace{1cm}\nonumber\\
&=&
P_1(t_1)
\prod_{i=2}^n \lambda(t_i) f(\Lambda(t_i)-\Lambda(t_{i-1});\lambda(t_i)^{2-\alpha}\phi)P_0((t_n,T]) .
\label{eq:likelihood2}
\end{eqnarray}
For spike trains consisting of $M$ repeated trials, $\{t_i^j\}_{j=1}^M:=\{t_1^j,\ldots,t_{n_j}^j\}_{j=1}^M$, 
the log likelihood function of $(\phi,\alpha)$, given $\{\lambda(t)\}$, is obtained as 
\begin{eqnarray}
L(\phi,\alpha;\{\lambda(t)\},\{t_i^j\}_{j=1}^M)
 &=& 
\sum_{j=1}^M \sum_{i=2}^{n_j}   \big\{ \log\lambda(t_i^j) + 
\log f(u_i^j;\xi_i^j)
\big\} \nonumber\\
& & { } + 
\sum_{j=1}^M \big\{ \log P_1(t_1^j) + \log P_0((t_n^j,T]) \big\},
\label{eq:loglikelihood}
\end{eqnarray}
where $\xi_i^j = \lambda(t_i^j)^{2-\alpha}\phi$ and $u_i^j = \Lambda(t_i^j)-\Lambda(t_{i-1}^j)$.
In the following analysis, we assume that there are many spikes in each trial $(n_j\gg1)$, so that the last two terms in Eq.~(\ref{eq:loglikelihood}) are negligible. 

If the firing rate $\lambda(t)$ is not known, an estimated firing rate $\hat{\lambda}(t)$ may be used, and the maximum likelihood estimator (MLE) $(\hat{\phi},\hat{\alpha})$ is obtained by maximizing Eq.~(\ref{eq:loglikelihood}) with respect to $(\phi,\alpha)$. 
The MLE does not generally admit closed form solutions, and is obtained by maximizing the log likelihood function numerically. 
In the numerical studies, we use a rectangular sliding window (\ref{eq:sample-mean}) to compute $\hat{\lambda}(t)$, and use a MATLAB function ``\texttt{fminsearch}" to maximize Eq.~(\ref{eq:loglikelihood}). 
We will discuss alternative methods for estimating $\lambda(t)$, and for estimating $(\phi,\alpha)$ together with $\lambda(t)$ rather than separately, in section~\ref{sec:discussion}. 

\begin{figure}[htbp]
\begin{center}
\includegraphics[width=13cm]{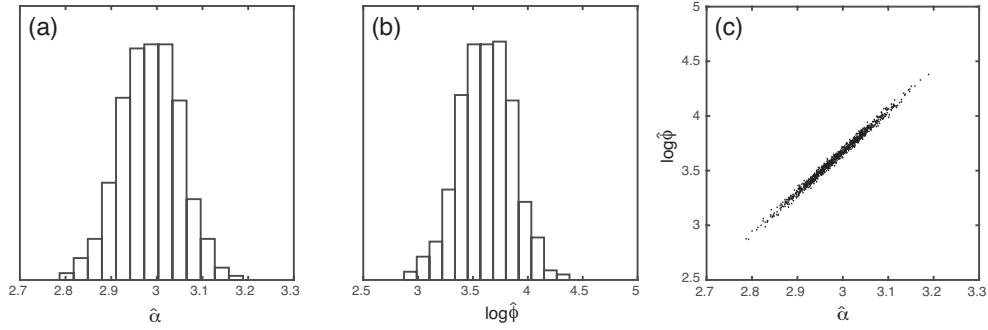}
\end{center}
\caption{
Histograms of $\hat{\alpha}$ (a) and $\log\hat{\phi}$ (b), obtained by $10^3$ repeated simulations, with $M=100$ spike trains. 
Both $\hat{\alpha}$ and $\log\hat{\phi}$ are approximately normally distributed. 
(c) presents a scatter plot of $(\hat{\alpha},\log\hat{\phi})$, showing that $\hat{\alpha}$ and $\log\hat{\phi}$ are linearly related.
}
\label{fig:siml_hist}
\end{figure}

The numerical studies in the following section show that the distribution of $\hat{\phi}$ is right-skewed because $\phi>0$, but that $\log\hat{\phi}$ is approximately normally distributed (figure~\ref{fig:siml_hist}b). 
Therefore, we consider the variance of $\hat{\eta} = \log\hat{\phi}$. 
By differentiating the log likelihood (\ref{eq:loglikelihood}) with respect to $\eta (=\log\phi)$ and $\alpha$, the observed information matrix is obtained as
\begin{eqnarray}
J(\eta,\alpha) &=&  
-\left(
\begin{array}{cc}
 \frac{\partial^2L}{\partial\eta^2} & \frac{\partial^2L}{\partial\eta\partial\alpha} \\
 \frac{\partial^2L}{\partial\eta\partial\alpha} &  \frac{\partial^2L}{\partial\alpha^2} 
\end{array}
\right) \nonumber\\
&=& 
-\sum_{j=1}^M\sum_{i=2}^{n_j} 
\Bigg\{
{\xi_i^j}^2
\frac{\partial^2}{\partial{\xi_i^j}^2}\log f(u_i^j;\xi_i^j) 
+
\xi_i^j
\frac{\partial}{\partial\xi_i^j}\log f(u_i^j;\xi_i^j)
\Bigg\} A_i^j,
\label{eq:obsinfo}
\end{eqnarray}
where 
\begin{equation}
A_i^j = 
\left(
\begin{array}{cc}
1 & -\log\lambda(t_i^j) \\
-\log\lambda(t_i^j) & \{\log\lambda(t_i^j) \}^2
\end{array}
\right) .
\label{eq:matrixA}
\end{equation}
Note that the rank of matrix $A_i^j$ is 1, but the rank of $J(\eta,\alpha)$ generally becomes 2.\footnote{
The rank of $J(\eta,\alpha)$ becomes 1 when the firing rate is constant.
}
Further, if the probability density function $f$ satisfies the regularity conditions that ensure asymptotic normality of parameter estimators \citep{Casella2002},  
the asymptotic variance matrix of the MLE is given by $J(\hat{\eta},\hat{\alpha})^{-1}$, with which
the confidence intervals for $\hat{\eta}$ and $\hat{\alpha}$ are constructed as 
\begin{equation}
\hat{\eta} \pm z \sqrt{ \big( J(\hat{\eta},\hat{\alpha})^{-1} \big)_{11} } 
\end{equation}
and 
\begin{equation}
\hat{\alpha} \pm z \sqrt{ \big( J(\hat{\eta},\hat{\alpha})^{-1} \big)_{22} } \ , 
\end{equation}
where $z$ is the critical value.

\subsubsection{Choice of ISI density function}
\label{sec:tweedie}

The ISI density function $f(x;\phi)$ is one of the building blocks of the proposed statistical model. 
Any ISI density function with a finite variance represents a generalized rate-modulated renewal process (\ref{eq:cif-ns}). 
Presently, we use a Tweedie distribution, a special case of an exponential dispersion model \citep{Jorgensen87,Jorgensen97}.
This is a two-parameter distribution, consisting of a linear exponential family with an additional dispersion parameter. These distributions play an important role in statistics, because they are the response distributions for generalized linear models \citep{McCullagh89}. 
A Tweedie distribution is an exponential dispersion model that has scale invariance (\ref{eq:scaling-pdf}), and includes probability distributions commonly used for describing the ISI variability, such as the gamma (for $\alpha=2$) and inverse Gaussian (for $\alpha=3$) distributions as special cases. These properties make a Tweedie distribution an obvious choice for $f(x;\phi)$.

Exponential dispersion models have a probability density function of the form,
\begin{equation}
f(x;\mu,\phi) = c(x,\phi)\exp\bigg[\frac{1}{\phi}\{x\theta-\kappa(\theta)\}\bigg],
\label{eq:edm}
\end{equation}
where $\theta$ is the canonical parameter, and $\kappa(\theta)$ is the cumulant function, with derivatives being the cumulants of the distribution. 
In particular, its mean and variance are given by $\mu=\dot{\kappa}(\theta)$ and $\sigma^2 = \phi\ddot{\kappa}(\theta)$, respectively.
The mapping from $\theta$ to the ISI mean $\mu$ is invertible, and is written $\ddot{\kappa}(\theta) = V(\mu)$ for a suitable function $V(\mu)$, called the variance function. 
A Tweedie distribution is identified by a particular choice of the variance function, as $V(\mu)=\mu^{\alpha}$.
By equating $\ddot{\kappa}(\theta)=d\mu/d\theta=\mu^{\alpha}$ and solving for $\mu$ and $\kappa$, $\theta$ and $\kappa$ are obtained as
\begin{equation}
\theta = 
\left\{
\begin{array}{ll}
\frac{\mu^{1-\alpha}-1}{1-\alpha} & \alpha\neq1 \\
\log\mu  & \alpha=1
\end{array}
\right.
,
\label{eq:tweedie-theta}
\end{equation}
and
\begin{equation}
\kappa(\theta) = 
\left\{
\begin{array}{ll}
\frac{\mu^{2-\alpha}-1}{2-\alpha} & \alpha\neq2 \\
\log\mu  & \alpha=2
\end{array}
\right.
,
\label{eq:tweedie-kappa}
\end{equation}
where we chose $\kappa(\theta)=0$ and $\mu=1$ at $\theta=0$, without loss of generality. 
The factor $c(x,\phi)$ in Eq.~(\ref{eq:edm}), which is determined by the normalization condition, does not have a closed form, except for in special cases. 
We compute it numerically, using series expansion and the Fourier inversion formula \citep{Dunn05,Dunn08}.

\section{Results} \label{sec:result}

In this section, we demonstrate with simulations that the proposed statistical model (\ref{eq:cif-ns}) describes spike trains that have a wide range of variability, characterized by $\alpha$ and $\phi$.
We illustrate on simulated and experimental data that our inference method is capable of estimating $\alpha$ and $\phi$ from rate-modulated spike trains.

\subsection{Simulation study}

\begin{figure}[htbp]
\begin{center}
\includegraphics[width=13cm]{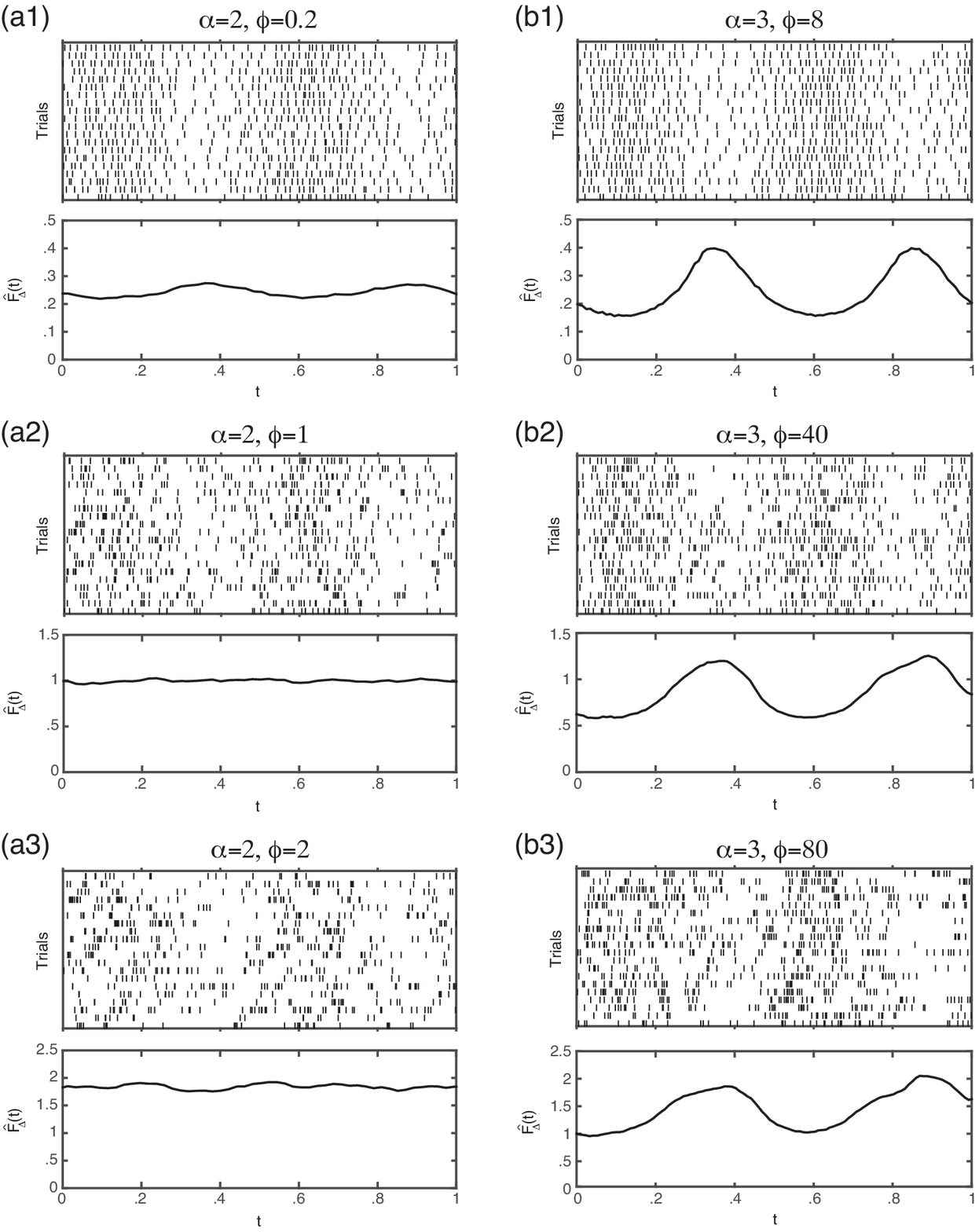}
\end{center}
\caption{
Raster plots of 20 spike trains simulated using Eqs.~(\ref{eq:probspike}) and (\ref{eq:frate}) and estimated Fano factor $\hat{F}_{\Delta}(t)$, for different $\alpha$ and $\phi$.
The Fano factor is almost constant for $\alpha=2$ (a1,2,3), while it is inversely related to the firing rate for $\alpha=3$ (b1,2,3). 
The overall Fano factor increases as $\phi$ increases, for fixed $\alpha$.
}
\label{fig:siml_fano}
\end{figure}

First, we simulate spike trains. 
The probability of a spike occurring in a short interval $(t,t+dt]$ is given by the conditional intensity function (\ref{eq:cif-ns}):
\begin{equation}
P\{N(t+dt)-N(t)=1; t_{N(t)},\{\lambda(t)\},\phi,\alpha\} = r(t;t_{N(t)},\{\lambda(t)\},\phi,\alpha)dt + o(dt).
\label{eq:probspike}
\end{equation}
Spike trains are simulated by discretizing the time into small bins ($dt=10^{-5}$), and evaluating Eq.~(\ref{eq:probspike}) in each bin. 
We use the firing rate function as 
\begin{equation}
\lambda(t) = 40 + 20\sin\frac{2\pi}{0.5}t,
\label{eq:frate}
\end{equation}
and generate $M$ spike trains in the time interval $t\in (0,1]$.
In order to compute the firing rate and the Fano factor, we use a sliding window of duration $\Delta=0.125$, in which an average of five spikes are expected to fall.
Let $N_{\Delta}^j(t)$ denote the number of spikes of the $j$th spike train in the counting window centered at $t$. 
The firing rate $\hat{\lambda}(t)$ and the Fano factor $\hat{F}_{\Delta}(t)$ in this window are computed, by averaging across trials, as  
\begin{equation}
\hat{\lambda}(t) = \frac{1}{M}\sum_{j=1}^M N_{\Delta}^j(t) \bigg/ \Delta,
\label{eq:sample-mean}
\end{equation}
and
\begin{equation}
\hat{F}_{\Delta}(t) = \frac{1}{M-1}\sum_{j=1}^{M}\{N_{\Delta}^j(t)-\hat{\lambda}(t)\Delta\}^2 \bigg/ \hat{\lambda}(t)\Delta.
\label{eq:sample-var}
\end{equation}
Figure~\ref{fig:siml_fano} displays the raster plots of 20 spike trains, and $\hat{F}_{\Delta}(t)$ computed with $M=10^4$ for different $\alpha$ and $\phi$.
We see that $\alpha$ and $\phi$ differentiate the variability of spike trains in different manners, as described in Eq.~(\ref{eq:fano-asymptotic}): $\phi$ scales the overall variability of spike trains, while $\alpha$ controls the dependency on the firing rate.
The Fano factor is inversely related to the firing rate for $\alpha=3$ (Figure~\ref{fig:siml_fano}b).
For $\alpha=2$, in which case the proposed model (\ref{eq:cif-ns}) corresponds to the ``conventional" rate-modulated renewal process (\ref{eq:cif-ns-cnv}), the Fano factor is almost constant, irrespective of the firing rate (Figure~\ref{fig:siml_fano}a).
For $\alpha=2$ and $\phi=1$, the spike trains become the inhomogeneous Poisson process (Figure~\ref{fig:siml_fano}a2).

\begin{figure}[htbp]
\begin{center}
\includegraphics[width=13cm]{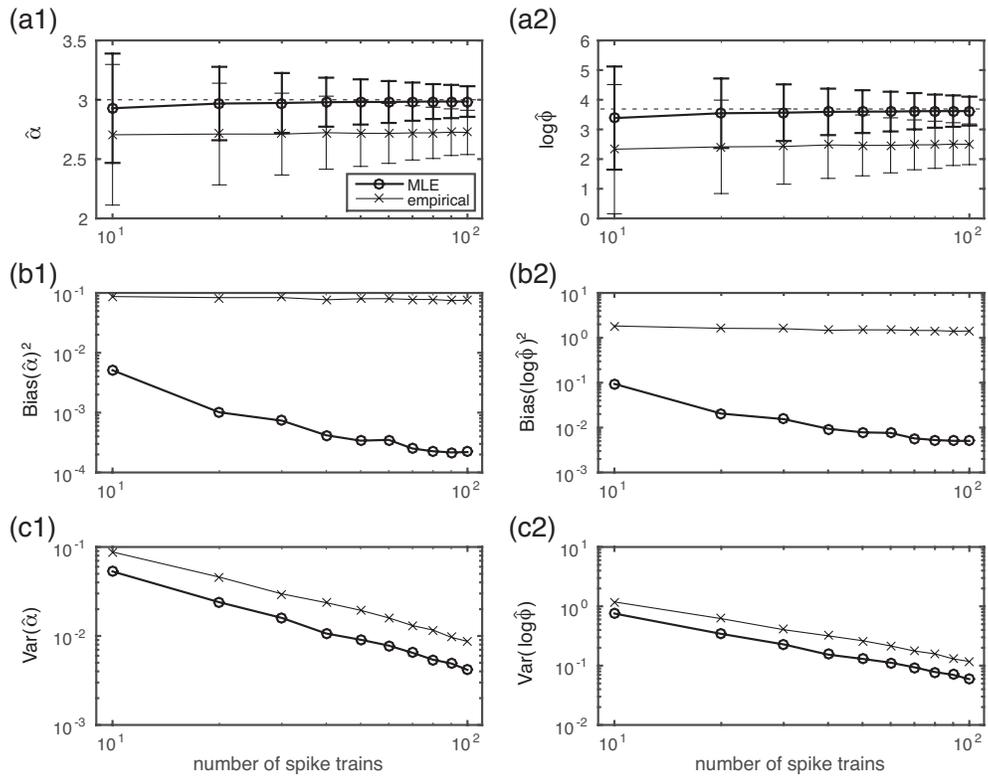}
\end{center}
\caption{
Estimates of $\alpha$ and $\log\phi$ as functions of the number of spike trains $M$. 
Results in this figure were computed by averaging across $10^3$ repeated simulations for each $M$.
Open circles and crosses represent the MLE and the empirical estimate, respectively. 
The true parameters are $\alpha=3$ and $\log\phi=\log40 (\approx3.69)$, represented by dashed lines in (a).
(a) presents the mean and 2SD of $\hat{\alpha}$ and $\log\hat{\phi}$. 
The errors in $\hat{\alpha}$ and $\log\hat{\phi}$ are decomposed into the squared bias (b) and variance (c). 
Both the bias and variance decrease as the number of spike trains $M$ increases.
The bias and variance of the MLEs are smaller than those of the empirical estimates.
}
\label{fig:siml_est}
\end{figure}

\begin{figure}[htbp]
\begin{center}
\includegraphics{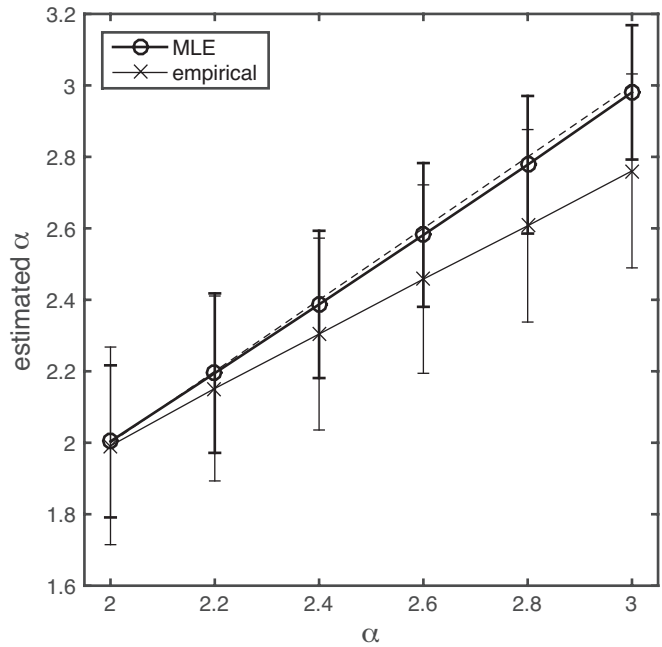}
\end{center}
\caption{
Plot of the MLE (open circles) and empirical estimate (crosses) of $\alpha$ against the true value, for $M=50$ spike trains. 
The mean and 2SD error bar were computed by averaging across $10^3$ repeated simulations. 
The dashed line represents the true value. 
The bias of the empirical estimate increases as $\alpha$ increases, while the mean of $\hat{\alpha}$ approximately matches the true value. 
}
\label{fig:siml_ealvsal}
\end{figure}

We simulated $M$ spike trains, from which the MLE  $(\hat{\alpha},\hat{\phi})$ was computed. 
We repeated the simulation $10^3$ times.
Figure~\ref{fig:siml_hist} shows that $\hat{\alpha}$ and $\log\hat{\phi}$ are approximately normally distributed, and that they are correlated. 
Figure~\ref{fig:siml_est}a plots $\hat{\alpha}$ and $\log\hat{\phi}$ against the number of spike trains $M$ (open circles). 
It is observed that $\hat{\alpha}$ and $\log\hat{\phi}$ converge to the true values as $M$ increases. 
The errors in $\hat{\alpha}$ and $\log\hat{\phi}$ are decomposed into the bias and variance, which are shown in figure~\ref{fig:siml_est}bc. 
Figure~\ref{fig:siml_ealvsal} displays the MLE $\hat{\alpha}$ of the exponent against the true value, ranging from $\alpha=2$ to $3$, where $\hat{\alpha}$ was computed with $M=50$ (open circles). 
The MLE approximately matches the true value in this range. 

For comparison, we computed an empirical estimate of ($\alpha, \log\phi$), using linear regression of $\{\log\hat{F}_{\Delta}(t)\}$ on $\{\log\hat{\lambda}(t)\}$ from Eqs.~(\ref{eq:fano-asymptotic}) and (\ref{eq:exponent}).
The results are plotted in figures~\ref{fig:siml_est} and \ref{fig:siml_ealvsal} (crosses). 
It is confirmed that the bias and variance of ($\hat{\alpha}, \log\hat{\phi}$) are smaller than those predicted by the empirical estimates.

\subsection{Experimental data}

We apply our method to two experimental datasets.
One dataset, labeled ``nsa2004.1",  is publicly available from the Neural Signal Archive \citep{Britten04}.
The spike data was recorded from 216 neurons in the visual cortical area MT of adult rhesus macaques. 
The recordings were obtained while a visual stimulus, consisting of a dynamic random dot pattern, was presented. 
Further experimental details can be found in \cite{Britten92}.
The other dataset, labeled ``ia-1",  is available from the CRCNS data sharing website \citep{Rokem09}.    
Spike trains were recorded from 43 auditory receptor cells of grasshoppers, while an auditory stimulus consisting of random amplitude modulations of wave was presented.
See \cite{Rokem06} for more details. 

\begin{figure}[htbp]
\begin{center}
\includegraphics{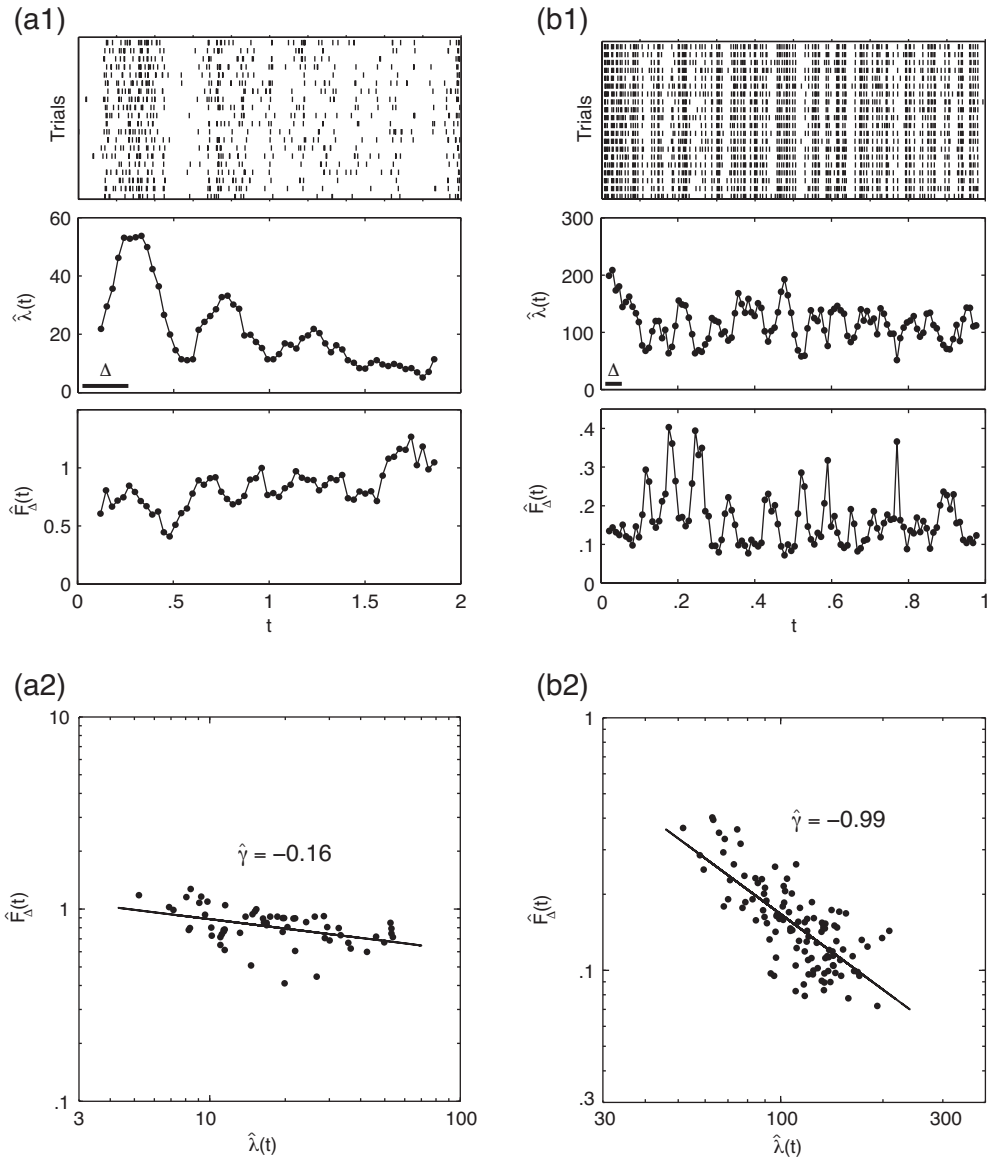}
\end{center}
\caption{
Representative sub-datasets for nsa2004.1 (a1) and for ia-1 (b1).
(Top) raster plot of 20 spike trains; 
(middle) estimated firing rate (the horizontal bar indicates the length of the counting window); 
(bottom) the Fano factor. 
The Fano factor is plotted against the firing rate on a log-log scale (a2 for nsa2004.1 and b2 for ia-1), on which linear regression was performed to obtain the exponent $\hat{\gamma}$. 
}
\label{fig:rdata_samples}
\end{figure}

Both datasets were divided into sub-datasets, consisting of multiple spike trains recorded from one cell under identical stimulus conditions. 
We selected sub-datasets containing $\ge 50$ trials, and with the mean firing rate $\ge 10$ spikes/s, due to the sufficiency of spikes for the analysis. 
Consequently, 193 sub-datasets for nsa2004.1 and 138 sub-datasets for ia-1 were used.
Representative sub-datasets for nsa2004.1 and ia-1 are shown in Figure~\ref{fig:rdata_samples}, together with the estimated firing rate $\hat{\lambda}(t)$ and Fano factor $\hat{F}_{\Delta}(t)$, computed with the sliding window whose length $\Delta$ was taken so that an average of five spikes are encompassed. 

\begin{figure}[htbp]
\begin{center}
\includegraphics[width=13cm]{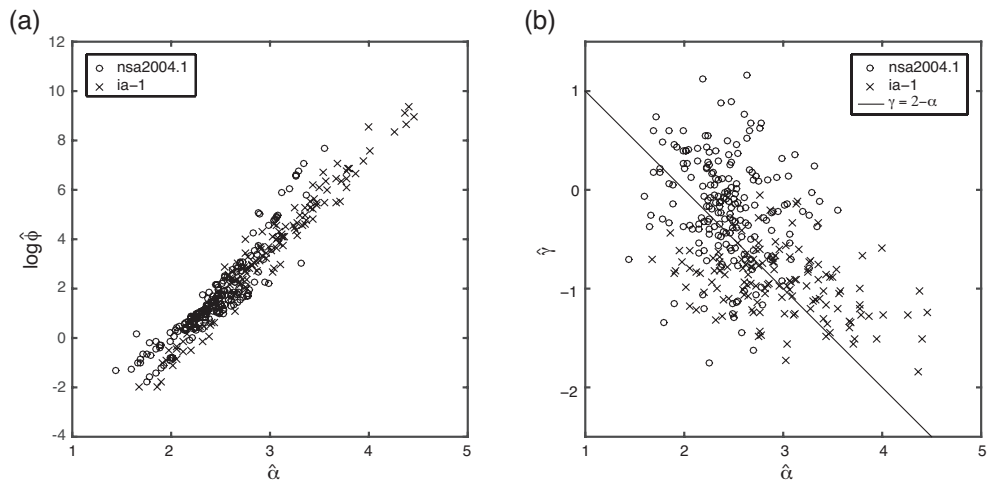}
\end{center}
\caption{
Scatter plot of ($\hat{\alpha},\log\hat{\phi}$) (a) and ($\hat{\alpha},\hat{\gamma}$) (b).
Open circles represent nsa2004.1 and crosses represent ia-1.
The solid line in (b) represents the scaling relation (\ref{eq:exponent}), around which the individual sub-datasets scatter.
}
\label{fig:rdata_result}
\end{figure}

For each sub-dataset, we obtained the MLE $(\hat{\alpha}, \hat{\phi})$. Figure~\ref{fig:rdata_result}a shows a scatter plot of $(\hat{\alpha}, \log\hat{\phi})$ (open circles stand for nsa2004.1 and crosses stand for ia-1).
The mean and standard deviations of the MLE are $\hat{\alpha}=2.43\pm0.38$ and $\log\hat{\phi}=1.52\pm1.69$ for nsa2004.1, and $\hat{\alpha}=2.96\pm0.58$ and $\log\hat{\phi}=3.37\pm2.46$ for ia-1. 
It is observed that a large portion of the $\hat{\alpha}$ are greater than two, and that on average $\hat{\alpha}$ of ia-1 is greater than that of nsa2004.1. 
This indicates that the firing variability tends to decrease as the firing rate increases, and that this tendency is stronger in ia-1 than in nsa2004.1.
In order to confirm this result, we estimated the exponent $\gamma$ of the Fano factor empirically for each sub-dataset, by performing linear regression of $\{\log\hat{F}_{\Delta}(t)\}$ on $\{\log\hat{\lambda}(t)\}$ (see Figure~\ref{fig:rdata_samples}a2,b2).
The estimated exponents, $\gamma$, are $\hat{\gamma} = -0.17 \pm 0.51$ for nsa2004.1 and $\hat{\gamma} = -0.94 \pm 0.33$ for ia-1. 
Figure~\ref{fig:rdata_result}b plots $\hat{\gamma}$ against $\hat{\alpha}$ in each of the sub-datasets, showing that the individual sub-datasets scatter around the line (\ref{eq:exponent}).

\section{Discussion} \label{sec:discussion}

This article was concerned the variability of spike trains, described by the power mean-variance relationship (\ref{eq:statlaw}). 
It was shown that this single assumption allows the spike trains to have a wide range of variability, characterized by $\phi$ and $\alpha$. 
By combining the power law with the time-rescaling transformation, we proposed generalized rate-modulated renewal processes, based on which a statistical method was developed for inferring $(\phi, \alpha)$ from rate-modulated spike trains. 

In our method, the firing rate $\lambda(t)$ was estimated separately from $(\phi, \alpha)$, using a rectangular counting window. 
We could use other methods, such as kernel density estimators or spline methods, which produce more precise rate estimates \citep{Kass03,Shimazaki10}. 
Alternatively, one may estimate $\lambda(t)$ together with $(\phi,\alpha)$, rather than separately. 
A principled method is to adopt a Bayesian framework, introducing a prior process of $\lambda(t)$ for regularization and computing the posterior process. 
Parameters of the prior process and $(\phi,\alpha)$ can be simultaneously optimized, by maximizing the marginal likelihood or the evidence \citep{Cunningham08,Koyama05,Koyama13}, which may improve the statistical efficiency.

It is often assumed that the variance of spike counts is proportional to their mean \citep{Averbeck09}, 
where the coefficient of proportionality (which corresponds to the Fano factor) may differ from unity due to a deviation from Poisson spiking. 
In our formulation, this assumption is relaxed, and we adopt one that the ratio of the count variance to the mean changes with the firing rate (Eq.~(\ref{eq:fano-asymptotic})), which is observed in a wide range of brain regions \citep{Kara00}.

The degree of irregularity of neural firing, which is measured by ISI statistics such as the {\it local variation} $L_V$ \citep{Shinomoto03}, is generally maintained {\it in vivo} cortical areas, while the firing rate varies in time \citep{Maimon09,Shinomoto09}.
This implies that the exponent of the power law (\ref{eq:scaling-isi}) in the ISI statistics is $\alpha\approx2$, from which a linear relationship between the mean and variance of spike counts ($\gamma\approx0$) is expected. 
On the other hand, steady discharges of X retinal ganglion cells, in response to stationary visual patterns, approximately obey the power law with $\alpha\approx3$ \citep{Troy92}, 
implying that a fixed ratio of the variance to the mean spike count no longer holds, but 
that the spike counts are less variable at higher rates \citep{Berry98,Reich98}. 

In the nervous system, neurons produce an action potential by integrating presynaptic inputs within tens of milliseconds, in which typically a few spikes come from each presynaptic neuron. 
This implies that the variance of spike counts in the integration time exhibits the power law, so that the presynaptic inputs have {\it signal-dependent noise}\footnote{
With a temporal resolution of this integration time, spike trains may be described as 
\begin{equation}
\frac{dN(t)}{dt} \approx \lambda(t) + \xi(t), \nonumber
\end{equation} 
where $\xi(t)$ is a white noise with $\mathrm{E}[\xi(t)]=0$ and 
$\mathrm{E}[\xi(t)\xi(s)]=\phi\lambda(t)^{\gamma+1}\delta(t-s)$.
}
that may be relevant to the computation carried out by the nervous system.
\cite{Ma06} hypothesized that the Poisson-like statistics in the responses of populations of cortical neurons may represent probability distributions over the stimulus, and implemented Bayesian inferences using linear combinations of the responses.
 A necessary condition in their hypothesis, which makes the Bayesian inferences possible, is that the variance of spike counts is proportional to the mean spike count ($\gamma=0$).
\cite{Lu13} showed that in controlling dynamical systems with noisy signals, precise control is achievable if the control signal has sub-Poisson noise ($\gamma<0$), while it is not achievable if the control signal has Poisson or supra-Poisson noise $(\gamma\ge0)$.

By analyzing a stochastic leaky integrate-and-fire model, we provided a possible mechanistic explanation for the origin of the power law (\ref{eq:statlaw}) with various exponents \citep{Koyama-FS}:
$\alpha=3$ may imply a supra-threshold firing regime, in which firing is driven by excitatory input; $\alpha=2$ may be interpreted as a sub-threshold firing regime, in which the membrane potential fluctuates below the threshold; 
and $\alpha=1$ may emerge when firing is strongly caused by large fluctuations of the membrane potential. 
Therefore, it is speculated that the ``intrinsic" exponent may reflect electrophysiological properties of individual cells or dynamical states of networks, and may vary across different brain areas. 
The proposed statistical framework offers a systematic way to explore the diversity of the variability of neural responses.

\appendix
\section{Derivation of the likelihood function}
\label{apdx:likelihood}

In this appendix, we derive Eq.~(\ref{eq:likelihood2}) from Eq.~(\ref{eq:likelihood1}). 
Using Eq.~(\ref{eq:cif-ns}), the second factor in the rhs of Eq.~(\ref{eq:likelihood1}) is rewritten as 
\begin{eqnarray}
\lefteqn{\prod_{i=2}^n r(t_i;t_{i-1},\{\lambda(t)\},\phi,\alpha)}\hspace{1cm}\nonumber\\
 &=& 
 \prod_{i=2}^n \lambda(t_i) f(\Lambda(t_i)-\Lambda(t_{i-1});\lambda(t_i)^{2-\alpha}\phi) \nonumber\\
 & & { } \times
\prod_{i=2}^n \Bigg[
 1-\int_{t_{i-1}}^{t_i}\lambda(v)f(\Lambda(v)-\Lambda(t_{i-1});\lambda(v)^{2-\alpha}\phi)dv
 \Bigg]^{-1}.
 \label{eq:stpdf2-proof1}
\end{eqnarray}
Taking the derivative of the logarithm of the last factor in Eq.~(\ref{eq:stpdf2-proof1}) leads to 
\begin{eqnarray}
\lefteqn{
\frac{d}{d t_i}
\log \Bigg[
 1-\int_{t_{i-1}}^{t_i}\lambda(v)f(\Lambda(v)-\Lambda(t_{i-1});\lambda(v)^{2-\alpha}\phi)dv
 \Bigg]
 }\hspace{1cm} \nonumber\\
 &=& 
-\frac{\lambda(t_i)f(\Lambda(t_i)-\Lambda(t_{i-1});\lambda(t_i)^{2-\alpha}\phi)}
{1-\int_{t_{i-1}}^{t_i}\lambda(v)f(\Lambda(v)-\Lambda(t_{i-1});\lambda(v)^{2-\alpha}\phi)dv}
\nonumber\\
&=&
- r(t_i;t_{i-1},\{\lambda(t)\},\phi,\alpha),
\end{eqnarray}
where the last equality comes from Eq.~(\ref{eq:cif-ns}).
Thus, we obtain 
\begin{eqnarray}
\lefteqn{
\prod_{i=2}^n\Bigg[1-\int_{t_{i-1}}^{t_i}\lambda(v)f(\Lambda(v)-\Lambda(t_{i-1});\lambda(v)^{2-\alpha}\phi)dv\Bigg]
}\hspace{1cm} \nonumber\\
&=& \exp\Bigg( -\int_{t_1}^{t_n}r(u;t_{N(u)},\{\lambda(t)\},\phi,\alpha)du \Bigg).
 \label{eq:stpdf2-proof2}
\end{eqnarray}
Substituting Eqs.~(\ref{eq:stpdf2-proof1}) and (\ref{eq:stpdf2-proof2}) into  Eq.~(\ref{eq:likelihood1}) leads to Eq.~(\ref{eq:likelihood2}).

\section*{Acknowledgments}

I would like to thank K. H. Britten, M. N. Shadlen, W. T. Newsome, and J. A. Movshon for uploading their experimental data to the Neural Signal Archive. 
I am also indebted to A. Rokem for collecting the data at the lab of A. Herz and providing it through the CRCNS program.
This research was supported by JSPS KAKENHI Grant Number 24700287.

\bibliographystyle{apalike}
\bibliography{NECO-09-14-2221-bibtex}

\newpage

\end{document}